# Deep Learning Models for Arrhythmia Classification Using Stacked Time-frequency Scalogram Images from ECG Signals

Parshuram N. Aarotale[1] and Ajita Rattani[2]

[1] Wichita State University, Kansas, USA
[2] University of North Texas at Denton, Texas, USA

**Abstract**

*Electrocardiograms (ECGs), a medical monitoring technology recording cardiac activity, are widely used for diagnosing cardiac arrhythmia. The diagnosis is based on the analysis of the deformation of the signal shapes due to irregular heart rates associated with heart diseases. Due to the infeasibility of manual examination of large volumes of ECG data, this paper aims to propose an automated AI-based system for ECG-based arrhythmia classification. To this front, a deep-learning-based solution has been proposed for ECG-based arrhythmia classification. Twelve lead electrocardiograms (ECG) of length* 10 *sec from* 45, 152 *individuals from Shaoxing People's Hospital (SPH) dataset from PhysioNet with four different types of arrhythmias were used. The sampling frequency utilized was* 500 *Hz. Median filtering was used to preprocess the ECG signals. For every* 1 *sec of ECG signal, the time-frequency (TF) scalogram was estimated and stacked row-wise to obtain a single image from 12 channels, resulting in* 10 *stacked TF scalograms for each ECG signal. These stacked TF scalograms are fed to the pretrained convolutional neural network (CNN), 1D CNN, and 1D CNN-LSTM (Long short-term memory) models, for arrhythmia classification. The fine-tuned CNN models obtained the best test accuracy of about* 98% *followed by* 95% *test accuracy by basic CNN-LSTM in arrhythmia classification.*

## 1. Introduction

Cardiovascular disorders are a leading cause of death and are highly prevalent affecting 49.5% of people over the age of 20 in the USA alone [1]. An electrocardiogram (ECG) examines the electrical activity of the heart and aids in the diagnosis of cardiovascular disorders by detecting different irregularities in patterns. ECG images have traditionally been manually evaluated by medical professionals to evaluate a patient's heart health. But as technology developed and due to infeasibility in the manual examination of large volumes of ECG data, a wide range of automated diagnostic tools appeared [2]. Early detection and an improved prognosis are the advantages of automatic cardiac abnormality identification using ECGs. Thus, improving the detection of arrhythmia and providing vital assistance to medical professionals [2]. An electrocardiogram (ECG) report is obtained using a Holter monitor and typically using a 12-lead design made up of three limb leads, three pressurized limb leads, and six thoracic leads [3]. While 12-lead ECGs are often employed in such applications, using fewer leads allows for inexpensive, portable, and user-friendly point-of-care equipment [4].

Apart from conventional machine learning techniques like feed-forward neural networks, decision trees, and support vector machines[5, 6], deep learning techniques have lately received significant attention in automatic cardiac abnormality identification using ECGs [7–10].

In this paper, we **propose** and assess the applicability of a deep learning-based solution for classifying 4 different classes of cardiac anomalies using stacked time-frequency scalogram images from 12-lead ECG data.

## 2. Related work

Wu et al. [11] proposed an efficient 1D-CNN framework consisting of 12 layers. Additionally, the authors implemented a denoising technique based on wavelet-based filtering. Niu et al. [3] proposed a novel adversarial deep-learning architecture for ECG classification. The authors enhanced the dataset size with the aid of domain adaptation for improved generalization. A new deep learning framework using a combination of CNN and bidirectional LSTM models was proposed by Liang et al. [10]. For ECG classification, Abdullah and Ani [12] proposed a combination of 1D CNN and long short-term memory (LSTM) based system. For the 10-class classification of arrhythmias, Abdalla et al. [13] used an 11-layer convolutional neural network (CNN) model. A bidirectional long-short-term memory network based on Recurrent Neural Network(RNN) was proposed by Kim and Pyun [14] for the classification of ECG signals.

Alarsan [9] evaluated multiple machine-learning meth-ods for the classification of heart rhythms. They used the MLlib, a machine learning library, to evaluate the machine learning models. The authors suggested that the random forest classifier obtained the highest accuracy. A hybrid CNN-LSTM model for categorizing arrhythmia was proposed by Liu et al. [15].

To effectively classify arrhythmia, Zhang et al. [16] used a 1D CNN model with a global average pooling layer. Convolutional neural networks were used by Mattila et al. [17] to classify arrhythmias for patients. For the ECG classification task, T Rahhal et al.[18] used a pre-trained VGGNet model. They transformed the signals into the time domain using a continuous wavelet transform. A bi-directional LSTM-based deep learning ECG classifier was proposed by Yildirim [19] for ECG classification.

## 3. Methods

### 3.1. Dataset

Twelve lead electrocardiograms (ECG) of length 10 sec from 45, 152 individuals from Shaoxing People's Hospital (SPH) dataset [20, 21] from PhysioNet with four different types of arrhythmias i.e., atrial fibrillation (AFIB), supraventricular tachycardia (ST), sinus bradycardia (SB), and sinus rhythm (SR) is used in this study. The sampling frequency utilized was 500 Hz for recording signals. The institutional review boards at Shaoxing People's Hospital and Ningbo First Hospital approved the study and granted the waiver request to acquire informed consent.

### 3.2. Time frequency representations of ECG signals

All ECG signals are pre-processed using median filtering. Then, for every 1 sec of ECG signal, the time-frequency (TF) scalogram was estimated, resulting in 10 TF scalograms for each ECG signal. TF scalograms are calculated from each channel and stacked row-wise to obtain a single image from 12 channels for 1 sec, as shown in Figure 1. These stacked time-frequency scalogram images are then fed to deep learning models such as ResNet50 [22], EfficientNetB0 [23], and MobilNet-V2[24] for ECG Arrhythmia classifications. Additionally, these images are fed to the CNN model developed from scratch with self-attention and multi-head attention mechanism for ECG Arrhythmia classification. For 1D CNN models, scalogram images are wrapped to a 1D feature vector.

### 3.3. Fine-tuned Deep Learning Models

This work utilizes different deep learning models such as ResNet50 [22], EffficientNetB0 [23] and Mobil-

Table 1. Architecture of 2D-CNN used in this study.

| Layer Type | Output Dimensions | Parameters |
|---|---|---|
| Conv | 223x223x32 | 2,432 |
| Max-Pool | 111x111x32 | 0 |
| Conv | 107x107x64 | 51,264 |
| Max-Pool | 53x53x64 | 0 |
| FC (with Dropout) | 500 | 9,012,500 |
| FC | 256 | 128,256 |
| FC | 64 | 16,448 |
| FC (Output) | 4 | 260 |

NetV2 [24] for ECG classification. These models are fine-tuned using TF scalogram images with a batch size of 32, a learning rate of 0.001, and using Adam optimizer [25]. These models are pre-trained ImageNet models and perform well for Arrhythmia classification after fine-tuning.

### 3.4. Basic Attention-based Convolutional Neural Network Architectures

For cross-comparison, we have used a simple CNN-based classification model. This model consists of two convolutional layers, followed by max-pooling layers; then feature maps are flattened into a single vector, which is then passed through a series of fully connected layers of size 500, 256, and 64 neurons. This is followed by a dropout layer (Dropout rate of 0.5) and the final classification layer of size 4 neurons, corresponding to four classes.

The self- and multi-head attention is applied to the basic CNN to obtain attention-weighted features. To this front, the input feature map is converted into query (Q), key (K), and value (V) representations through independent convolutional layers, with Q and K being reduced to 1/8th of the input channels. Q and K are then used to compute attention weights, to prioritize relevant spatial areas. For 1D-CNN-LSTM architecture, the scalogram images are resized to a 1D vector for input to the CNN model. The convolutional layers are followed by a three-layer LSTM of size 128 hidden units, followed by a dropout and the final classification layer of 4 neurons.

### 3.5. Performance Metrics

The standard performance evaluation metrics i.e., precision, recall, accuracy, and specificity [26, 27], were employed to assess the performance of the deep-learning models for arrhythmia classification.

## 4. Results and Discussion

Table 2 shows the precision, recall, accuracy, and specificity of different deep-learning models for arrhythmia classification. It can be seen that the best performance is obtained by the fine-tuned ResNet-50 model with precision, recall, specificity, and accuracy of 0.96, 0.96, 0.99, and 98.29, respectively. Overall, the fine-tuned version of the pre-trained models obtained better performance than the CNN models trained from scratch. Among the CNN models trained from scratch, 1D-CNN+LSTM obtained the best performance with precision, recall, specificity, and accuracy of 0.90, 0.89, 0.96, and 95.07, respectively. The least performance is obtained by the basic 2D CNN model without an attention mechanism.

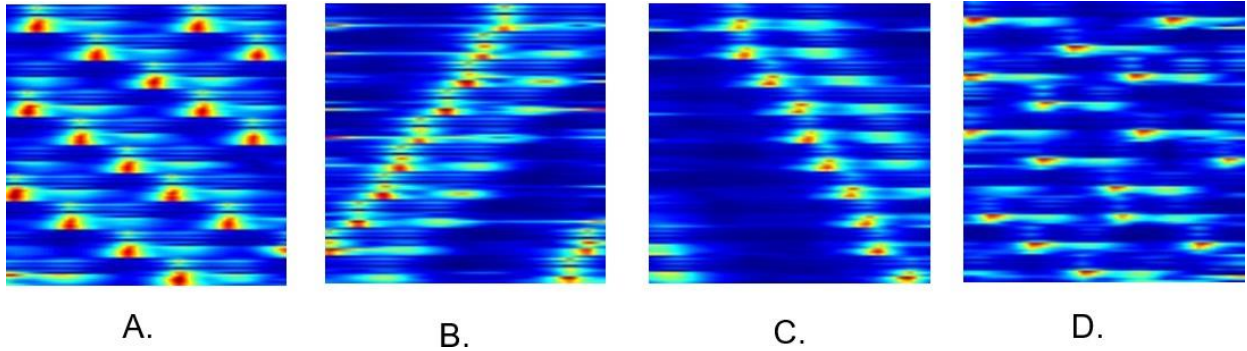

Figure 1. Time-frequency Scalogram stacked row-wise to obtain a single image from a 12 channel sample ECG signal for the four arrhythmia classes, A. supraventricular tachycardia (ST), B. sinus rhythm (SR), C. sinus bradycardia (SB), and atrial fibrillation (AFIB), respectively.

Table 2. Comparison of Deep-learning Model for ECG-based Arrhythmia Classification using Accuracy (A), Precision (P), Recall (P) and Specificity (S) when evaluated on the SPH test set.

| Model | A(%) | P | R | S |
|---|---|---|---|---|
| ResNet 50 [22] | 98.29 | 0.96 | 0.96 | 0.99 |
| EfficientNet-B0 [23] | 98.26 | 0.98 | 0.98 | 0.99 |
| MobileNet-V2[24] | 98.23 | 0.97 | 0.97 | 0.99 |
| 1D-CNN | 92.78 | 0.84 | 0.83 | 0.94 |
| 1D-CNN-LSTM | 95.07 | 0.90 | 0.89 | 0.97 |
| Simple CNN | 89.76 | 0.80 | 0.79 | 0.93 |
| CNN-SA | 92.48 | 0.81 | 0.81 | 0.94 |
| CNN-MHA | 91.52 | 0.79 | 0.78 | 0.93 |

A similar study in [28] used the SPH dataset and extracted the TF scalogram images from the ECG signals, separately for each channel, for training the Squeeze-Net model. The model obtained an accuracy score of 0.36 for arrhythmia classification using ECG signals. This suggests the efficacy of using stacked Scalogram images from all the leads together (obtaining around 98.29% accuracy) using our study for ECG-based arrhythmia classification.

## 5. Conclusions

In this study, we evaluated the efficacy of stacked time-frequency scalogram images from 12-lead ECG data for arrhythmia classification. Experimental investigation on the SPH dataset using different deep-learning architectures suggested the efficacy of the fine-tuned CNN models pretrained on ImageNet weights for arrhythmia classification. On cross-comparison with existing studies on deep-learning models based on a single TF scalogram image for each lead, stacked TF-based scalogram along with the fine-tuned SOTA CNN obtained a significant performance improvement. As a part of future work, cross-dataset evaluation of the deep-learning models trained on the SPH dataset will be conducted. Further, the evaluation will be conducted on an extended list of arrhythmia classes using advanced deep-learning models based on self-supervised learning.

Address for correspondence:

Dr. Ajita Rattani
Dept of Computer Science and Engineering,
University of North Texas at Denton, USA.
ajita.rattani@unt.edu